\documentclass[%
iop,
 amsmath,amssymb,
 reprint,%
]{revtex4-2}
\usepackage{graphicx}
\usepackage{dcolumn}
\usepackage{bm}%
\usepackage[colorlinks=true,citecolor=blue]{hyperref}
\hypersetup{colorlinks=true,citecolor=blue,linkcolor=red,urlcolor=blue}

\usepackage{float}
\usepackage{amsmath}

\usepackage{color}
\definecolor{Green}{RGB}{0,204,102}
\definecolor{Purple}{RGB}{102,0,255}
\definecolor{Blue}{RGB}{51,153,255}
\definecolor{Red}{RGB}{151,010,010}

\makeatletter
\def\@bibdataout@aps{%
\immediate\write\@bibdataout{%
@CONTROL{%
apsrev41Control%
\longbibliography@sw{%
    ,author="08",editor="1",pages="1",title="0",year="1"%
    }{%
    ,author="08",editor="1",pages="1",title="",year="1"%
    }%
  }%
}%
\if@filesw \immediate \write \@auxout {\string \citation {apsrev41Control}}\fi 
}
\makeatother

\begin{document}

\title{Spin-orbit torques due to topological insulator surface states: an in-plane magnetization as a probe of extrinsic spin-orbit scattering}

\author{Mohsen Farokhnezhad}
\affiliation{Department of Physics, School of Science, Shiraz University, Shiraz 71946-84795, Iran}

\affiliation{School of Nanoscience, Institute for Research in Fundamental Sciences, IPM, Tehran, 19395-5531, Iran}

\affiliation{School of Physics, Institute for Research in Fundamental Sciences, IPM, Tehran, 19395-5531, Iran}

\author{James H. Cullen}
\email[Corresponding author email: ]{james.cullen@unsw.edu.au}
\affiliation{School  of  Physics,  University  of  New  South  Wales,  Kensington,  NSW  2052,  Australia}

\author{Dimitrie Culcer}
\affiliation{School  of  Physics,  University  of  New  South  Wales,  Kensington,  NSW  2052,  Australia}

\date{\today}

\begin{abstract}
Topological insulator (TI) surface states exert strong spin-orbit torques. When the magnetization is in the plane its interaction with the TI conduction electrons is non-trivial, and is influenced by extrinsic spin-orbit scattering. This is expected to be strong in TIs but is difficult to calculate and to measure unambiguously. Here we show that extrinsic spin-orbit scattering sizably renormalizes the surface state spin-orbit torque resulting in a strong density dependence. 
{ The magnitude of the renormalization of the spin torque and the effect of spin-orbit scattering on the relative sizes of the in-plane and out-of-plane field-like torques have strong implications for experiment: We propose two separate experimental signatures for the measurement of its presence.}
\end{abstract}
\maketitle

\section{Introduction}
The advent of topological insulators (TIs) has had a profound effect on condensed matter research in the past decade \cite{PhysRevLett.98.106803,PhysRevB.76.045302,RevModPhys.82.3045,moore2010birth, shen2012topological, franz2013topological,Chang2015}. Their chiral surface states have reformulated band theory, offering insights into topological ordering, and advancing quantum information and quantum computing~\cite{RevModPhys.82.3045, tian2017property}. Moreover, TIs provide a pathway towards efficient electrical control of magnetization in hybrid magnetic/TI systems~\cite{mellnik2014spin,fan2014magnetization,PhysRevLett.114.257202,PhysRevLett.119.077702,Fan2014,Wang2017,Li2019,liu2021,fukami2016magnetization,culcer2009,Ramaswamy2019,pan2022efficient}. Applying an in-plane electric field to the TI will generate a spin polarization in the chiral surface states via the Rashba-Edelstein effect (REE)~\cite{RevModPhys.76.323, RevModPhys.87.1213}. This spin polarization can exert a torque on the local magnetization of the magnetic layer, this is referred to as a spin-orbit torque (SOT). Spin-orbit torques are a powerful mechanism for manipulating magnetic spin textures~\cite{dieny2020opportunities, shi2019all, macneill2017control, kent2015new}. In particular current-induced spin-orbit torques have opened up the possibility for efficient electrical manipulation of spintronic devices~\cite{PhysRevB.66.014407, PhysRevB.80.134403, PhysRevB.86.014416, RevModPhys.91.035004, chernyshov2009evidence,ryu2013chiral, miron2011perpendicular, wadley2016electrical, bandyopadhyay2008introduction,Ramaswamy2018,Shao2021RoadmapOS,Nikolicnikolic2018,zhou2021modulation}, such as magnetic memory. SOTs have been explored in FM/heavy metal bilayers~\cite{miron2011fast,skinner2015complementary,PhysRevB.75.155323}, and it has been found that FM/TI structures can exhibit substantially greater SOTs.

\begin{figure}[t]
\centering
\includegraphics[scale=0.7]{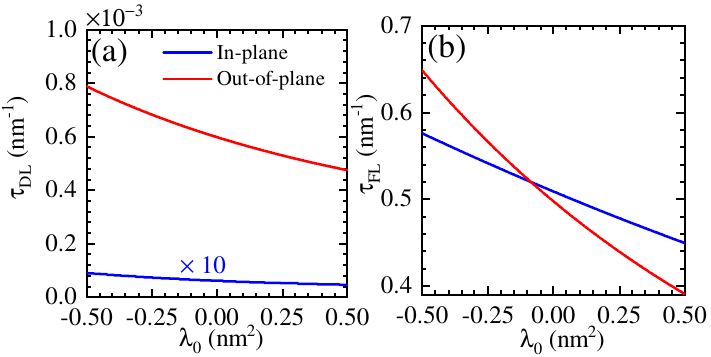}
\caption{The strength of the (a) damping-like and (b) field-like SOTs are plotted as a function of the extrinsic spin-orbit scattering strengths for both out-of-plane and in-plane magnetizations. The parameters used in the calculations are: Fermi energy $\varepsilon_{\rm F} = 200$ meV, magnetization $M = 10$ meV, impurity density $n_i = 10^{10}$ cm$^{-2}$, and $u_0 = 5$ eV nm$^2$. For the in-plane magnetization case, we set the warping term as $\lambda_w = 250$ meV nm$^{3}$, and both the azimuthal magnetization angle $\phi_m$ and electric field direction $\phi_E$ are set to $\pi/4$.}
\label{compare}
\end{figure}

The Rashba-Edelstein effect due to the surface states is expected to be the dominant contribution to TI spin torques~\cite{Ghosh2018,cullen2023spin}. In an out-of-plane magnetization the REE contribution to the spin-orbit torque is well understood~\cite{culcer2020transport,Kurebayashi2019}. However, the same effect in the surface states with an in-plane magnetization has not been studied as extensively. The case where the magnetization is in-plane is remarkably different from the out-of-plane case, as the magnetization term $\boldsymbol{M}_{\parallel}\cdot\boldsymbol{\sigma}_{\parallel}$ can be gauged away from the electron Hamiltonian, unless warping or extrinsic spin-orbit scattering are present. The first case has been studied in Ref.~\onlinecite{PhysRevB.99.155139}: in this work spin-orbit torques generated via intrinsic mechanisms and due to scattering from scalar impurity potentials were considered. The case of extrinsic spin-orbit scattering, where spin-orbit terms appear in the scattering potential itself, producing a random spin-orbit field that leads to spin-flip scattering, has not been studied. Yet spin-momentum locking in TIs ensures spin-dependent effects are inseparable from charge effects, and extrinsic spin-orbit scattering mechanisms are known to affect transport features such as weak antilocalization~\cite{adroguer2015conductivity}, hence they are expected very important for the surface state spin-orbit torque.

In light of this, we study electrically induced spin polarizations in magnetized surface states of a topological insulator, with an in-plane magnetization and extrinsic spin-orbit scattering, and the induced spin torque. We calculate the linear response of the density matrix, using a quantum kinetic equation based on the quantum Liouville equation~\cite{Culcer_2017, PhysRevResearch.4.013001, farokhnezhad2022spin}. We consider both scalar and spin-orbit coupled impurities and we solve the kinetic equation up to second order in the spin-orbit scattering parameter. Spin-orbit coupled scattering potentials cause electrons with different spins to be deflected in different directions via skew scattering~\cite{d1971possibility, PhysRevLett.83.1834, PhysRevLett.95.166605} and side-jump scattering~\cite{PhysRevB.2.4559}. The non-equilibrium density matrix obtained from the kinetic equation is then used to find electrically induced spin polarizations. The spin torque is then extracted by calculating $(2M/\hbar){\bf m}\times\langle{{\bf s}}\rangle$. 

Our main findings are: (i) The Edelstein effect, the primary driver of the surface state spin torque, is strongly renormalized by extrinsic SO scattering. Thus, while the in-plane magnetization has no effect on conduction electrons without spin-orbit scattering, the conduction electrons have a strong effect on the magnetization through the spin-orbit torque, and extrinsic spin-orbit scattering affects the spin-orbit torque both quantitatively and qualitatively. Most importantly, the field-like torque for the case with an out-of-plane magnetization will have a greater dependence on the spin-orbit scattering parameter than the case with an in-plane magnetization. This is demonstrated in Fig.~\ref{compare}. Since the relative sizes of the in-plane and out-of-plane field-like torques depend critically on extrinsic spin-orbit scattering and are strong functions of density, this provides a potential way to detect the presence of extrinsic spin-orbit scattering experimentally. This is important since the spin-orbit scattering parameter $\lambda$ cannot be calculated for topological insulators using the standard method for semiconductors, being a surface state parameter. { Similarly, we find that the spin-orbit scattering parameter significantly changes the Fermi energy dependence of the torque, providing another experimental signature of $\lambda$.} There are also no existing proposals for detecting its presence experimentally. Hence a spin torque measurement could provide a way out of this impasse. (ii) Novel spin polarizations arise from the interplay between the hexagonal warping, in-plane magnetization and spin-orbit scattering. However, it should be noted that the magnitudes of these effects are small. This SOT due to these polarizations will be most important when the spin density due to the Rashba-Edelstein effect is parallel to the magnetization, so that it does not contribute to the SOT.

\section{Model Hamiltonian and Theory}
We consider the surface states of a 3D TI in the presence of a magnetization ${\bf M}=M(\cos{{\phi}_m}, \sin{{\phi}_m})=M(m_x,m_y)$ where $M$ is the strength of exchange field, the magnetization is uniform and can point along any direction in 2D plane. The system is studied at zero temperature and thus we ignore thermal for the charge-spin conversion process. The exchange coupling between the magnetic moments and the conduction electron spins is given by the Zeeman term ${\bf M}\cdot {\bf \sigma}$ where ${\bf \sigma}$ is the vector of Pauli matrices representing the spin operators of conduction electrons. An in-plane magnetization leads to a shift in Dirac cone along the $k_x$ and $k_y$ directions as $\bar{\bf{k}}={\bf k}-{\bf k}_{m}$ with ${\bf k}_m=M(m_y, -m_x)/(\hbar{v_{\rm F}})$, while an out-of-plane magnetization opens a gap in the surface states~\cite{PhysRevLett.104.146802, PhysRevB.81.241410, PhysRevB.96.014408}.
The band Hamiltonian that describes low-energy excitations in the surface of 3D TIs incorporating the warping is given by~\cite{Liu2010,Hasan2010,PhysRevLett.103.266801}
\begin{eqnarray}
{\cal H}_0=\hbar v_{\rm F}({\hat y}\cdot {\bar{\bf{k}}}\sigma_{x}-{\hat x}\cdot {\bar{\bf{k}}}\sigma_{y})
+\lambda_w k^3_w \cos(3\varphi) \sigma_z
\end{eqnarray}
where $v_{\rm F}$ is the effective Fermi velocity,  $k_{w}^2=\bar{k}^2+k^2_{m}+2\bar{k}{k_m}\sin(\phi_m-{\theta}_{\bar{\bf k}})$ with $\theta_{\bar{\bf k}}=\tan^{-1}(\bar{k}_y/\bar{k}_x)$, ${\varphi}=\tan^{-1}(\frac{\bar{k}_y-m_x{k_m}}{\bar{k}_x+m_y{k_m}})$ and ${\bf k}$ is the momentum vector of electrons. The second term encapsulates the hexagonal warping, at low-Fermi energies, less than $0.55\varepsilon^*\approx 0.15$ eV, the impact of warping will be minimal.\cite{PhysRevLett.103.266801} However, it is necessary to consider it as it is important for the damping-like torque generated by the surface states. 

The eigenstates of the band Hamiltonian ${\cal H}_0$ are
\begin{eqnarray}
\vert u^{s}_{\bar{\bf k}} \rangle=\frac{1}{\sqrt{2}}
\begin{pmatrix}
s{i}\sqrt{1+s{b_{\bar{k}}}} e^{-i{\theta_{\bar{\bf k}}}} \\
\sqrt{1-s{b_{\bar{k}}}}
\end{pmatrix},
\end{eqnarray}
where $b_{\bar{k}}=\lambda_{w}k^3_{w}\cos(3\varphi)/\varepsilon_{\bar{\bf k}}$ and $s=\pm 1$. These energies of these eigenstates are $\varepsilon_{\bar{\bf k}}=\pm\sqrt{({\hbar}{v_{\rm F}}{\bar{k}})^2+\lambda^2_{w}k^6_{w}\cos^2(3\varphi)}$.

The total Hamiltonian describing the conduction electrons is
\begin{equation}\label{eq:Eq.1a}
{\cal H}={\cal H}_0+V({\bf r})+U(\bf r),
\end{equation}
where $V(\bf r)$ represents the electrostatic potential which has the form $V(\bf r)=e{\bf E}\cdot\bf r$ and $U=U(\bf r)$ represents the disorder scattering potential for the conduction electrons.  Here we consider the electric field ${\bf E}$ to be uniform. 
We consider short-range scalar impurities together with spin-orbit coupled impurities. For short-range scalar scattering, we consider $U_0({\bf r})=\sum_{i}{u_0}\delta({{\bf r}}-{{\bf R}}_i)$, where ${\bf R}_i$ indicate the random locations of the impurities and $u_0$ is a parameter that measures the strength of the disorder potential~\cite{PhysRevResearch.4.013001}. 
The extrinsic spin-orbit scattering~\cite{sinitsyn2007semiclassical} is given by $U_{{so}}({\bf r}) = \lambda_0{{\bf \sigma}}\cdot{{\bf \nabla}} U_0({\bf r}) \times{\bf p}$, where ${\bf p}$ is the momentum operator and $\lambda_0$ is the effective strength of the spin-orbit in the impurity potential. The extrinsic SO scattering term can be likened to a random effective magnetic field, which depends on an electron’s incident and scattered wave vectors. $\lambda_0$ can have either sign and it is assumed that $|\lambda_0| < 1$. Furthermore, we assume $\lambda_{w}{\bar k}_{\rm F}^2/(\hbar{v_{\rm F}})\ll{1}$ and $k_{m}/{\bar k}_{\rm F}\ll{1}$ in all our calculations. Subsequently, the disorder potential including scalar and SO impurities in the reciprocal space reads as

\begin{eqnarray}\label{eq:Eq. (219)}
U^{ss'}_{\bar{\bf k}\bar{\bf k}'}={u_0}\langle{{u}^{s}_{\bar{\bf k}}}|\Big[\sigma_0+i{\lambda}_{1}{\sigma}_z{\sin{{\gamma}_1}}-2i{\lambda}_{2}{\sigma}_z{\sin{\frac{{\gamma}_1}{2}}\sin{\frac{{\gamma}_2}{2}}}\Big]|{{u}^{s'}_{\bar{\bf k'}}}\rangle,\nonumber
\end{eqnarray}
where $\lambda_{1}=\lambda_{0}{\bar{k}_{\rm F}^2}$, $\lambda_{2}=\lambda_{0}{\bar{k}_{\rm F}}{k_m}$ with $k_m=M/({\hbar{v_{\rm F}}})$, ${\gamma}_{1} = \theta_{\bar{\bf k'}}-\theta_{\bar{\bf k}}$ and ${\gamma}_{2} = \theta_{\bar{\bf k'}}+\theta_{\bar{\bf k}}-2\phi_{m}$. For the sake of simplicity, we substitute ${\bar k}$ with $k$ from now on. Averaging over impurity configurations, the first-order term in the potential vanishes due to the randomness of the impurity locations, while the disorder averaged second-order term is
\begin{equation}
\langle{U}^{mm'}_{{\bf k{\bf k'}}}{U}^{m''{m'''}}_{{\bf k'{\bf k}}}\rangle={n_i}{U}^{mm'}_{{\bf k{\bf k'}}}{U}^{m''m'''}_{{\bf k'{\bf k}}},
\end{equation}
where $n_i$ is the impurity concentration. Using Fermi's golden rule, the relaxation time $\tau_{tr}$ with index $m$ is calculated in the presence of impurities as
\begin{equation}
\frac{1}{\tau_{tr}(\bf k)}=\frac{2\pi}{\hbar}\sum_{m'}\int\frac{k'{dk'}{d{\theta'}}}{(2\pi)^2}\langle{U}^{mm'}_{{\bf k{\bf k'}}}{U}^{m'{m}}_{{\bf k'{\bf k}}}\rangle{\delta(\varepsilon^{m}_{\bf k}-\varepsilon^{m'}_{\bf k'})}.
\end{equation}

The relaxation time for both spin states is given by 
\begin{eqnarray}\label{tau1} 
\frac{1}{\tau_{tr}(k_{\rm F})}=\frac{1}{\tau}\Big[1+\lambda_{1}+\frac{\lambda^2_1}{2}+\frac{3}{2}\lambda^{2}_2+\frac{3}{4}\alpha^2_{w}\Big].
\end{eqnarray}
where $\alpha_w=\lambda_{w}k_{\rm F}^3/ \epsilon_{\rm F}$, ${\tau^{-1}}={n_i{u^2_0}\pi\rho(k_{\rm F})}/{\hbar}$, $\rho(k_{\rm F})={k_{\rm F}}/{2\pi\hbar{v_{\rm F}}}$, $\epsilon_{\rm F}=\hbar v_{\rm F} k_{\rm F}$ distinct from $\varepsilon_{{\bf k}_{\rm F}}$, and the Fermi momentum $k_{\rm F} \approx \frac{\varepsilon_{\rm F}}{\hbar{v_{\rm F}}}[1-\frac{1}{4}(\frac{\lambda_{w}}{\hbar{v_{\rm F}}})^2(\frac{\varepsilon_{\rm F}}{\hbar{v_{\rm F}}})^4)]-k_m$.

\section{Quantum Kinetic Theory and Spin-Orbit Torque}
The density matrix $\rho=f_{\bf k}+g_0$ characterizes the system considered in Eq. (\ref{eq:Eq.1a}), where $f_{\bf k}$ is averaged over disorder configurations and has matrix elements connecting different bands, and $g_0$ is the fluctuating part.
For the disorder-averaged component, the density matrix obeys the quantum kinetic equation~\cite{Culcer_2017}
 \begin{equation}\label{eq:QKE}
    \frac{\partial f_{\bf k}}{\partial t} + \frac{i}{\hbar} \, [{\cal H}_{0{\bf k}}, f_{\bf k}] + J_{0}(f_{\bf k}) = \frac{e{\bf E}}{\hbar}\cdot\frac{Df_{\bf k}}{D{\bf k}} - J_E(f_{\bf k}),
\end{equation}
where the covariant derivative is $\frac{Df_{\bf k}}{D{\bf k}} = \frac{\partial f_{\bf k}}{\partial{\bf k}} -i[{\bf \mathcal{R}}_{\bf k}, f_{\bf k}]$ and ${\bf \mathcal {R}} ^{ss'}_{\bf k}=i\langle {{u}^{s}_{\bf k}}|\nabla_{\bf k}{{u}^{s'}_ {\bf k }}\rangle$ is the Berry connection.  $J_0(f_{\bf k})$ is the collision integral owing to the impurity potential. The electric field in the Hamiltonian induces a correction to the Green's function and through it a correction to the collision integral, $J_E(f_{\bf k})$, which must be considered~\cite{PhysRevResearch.4.013001,farokhnezhad2022spin}.

{While the focus of this work is not on the technical aspect of the calculation, we would like to note that the treatment of the electric field correction to the scattering integral while including spin-orbit coupled impurity potentials is unique to to our approach~\cite{farokhnezhad2022spin}. Furthermore, calculating extrinsic spin-orbit scattering effects in a TI surface state Hamiltonian with an in-plane magnetization, in which the magnetization can be gauged out of the Hamiltonian but must be considered in the impurity potential, is another novel technical aspect of this work.}

To calculate the spin polarization we first calculate the matrix elements of the density matrix, then we trace it with the spin operator. More detailed calculations can be found in Ref.~\cite{Note2}. After performing the necessary calculations, the leading order spin density can be expressed as follows:
\begin{equation}\label{leading1} 
{\langle}{\bf s}{\rangle}^{\text{leading}}=e\tau Z(k_{\rm F})\left[(4+5\alpha^2_{w})(\hat{\bf z}\times {\bf E})-\frac{3}{2}\alpha^3_{w}\Lambda\left(\frac{M}{\epsilon_{\rm F}}\right)\hat{\bf z}\right]
\end{equation}
where ${\langle}{\bf s}{\rangle}^{\text{leading}}$ is proportional to $\tau$, $Z(k_{\rm F})={v_{\rm F}}\rho(k_{\rm F})/[4(1+\lambda_1)+2\lambda^2_{1}+6\lambda^2_{2}+3\alpha^2_{w}]$ and 
\begin{align}\label{lambda1}
&\Lambda(Q)=Q^2\Big(163(m^2_{x}-m^2_{y}){E_x}+26{m_x}m_y{E_y}\Big)\nonumber\\
&+Q^4\Big(765(m^2_{x}-m^2_{y}){E_x}+1110{m_x}m_y{E_y}\Big).
\end{align} 

The presence of the warping term leads to an increase in the in-plane extrinsic spin density, primarily due to the strengthening of the SO scattering. We have performed calculations of the sub leading order spin density including calculations of skew scattering and side jump, we find in general these spin densities to be of negligible magnitude and will only be relevant in specific systems. Further details of these calculations can be found in the supplement~\cite{Note2}.
 
After calculating the spin density, we calculated the SOT. The spin-orbit torque is defined as $(2M/\hbar){\bf m}\times\langle{{\bf s}}\rangle$, where ${\bf m}$ represents the unit vector of the in-plane magnetization. 
The SOT can be decomposed into two parts based on their symmetry properties with respect to magnetization reversal: the field-like torque ${\bf m}\times{\bf \sigma}$ and the damping-like torque ${\bf m}\times({\bf m}\times{\bf \sigma})$, where ${\bf \sigma}$ is a vector independent of the magnetization. These components have distinct physical origins and effects on the magnetization dynamics. In the macrospin approximation, the field-like torque causes precession of the magnetization and, the damping-like torque damps this precession and causes the magnetization to align with ${\bf \sigma}$.

The leading order spin density eqn. (\ref{leading1}) will give rise to a field-like torque. From these terms it appears that for the field-like torque this vector ${\bf \sigma}$ will be almost exactly aligned along ${\bf\hat{z}}\times {\bf E}$ being rotated slighty toward $\hat{z}$ by the warping term, however, this is not exactly correct as the sub leading order terms of the spin density will cause a small azimuthal rotation of this vector. However, in our estimates this torque is 4 orders of magnitude smaller than the maximal torque due to the dominant leading order term $\propto{\bf\hat{z}}\times {\bf E}$.

\begin{figure}
\centering
\includegraphics[scale=1.2]{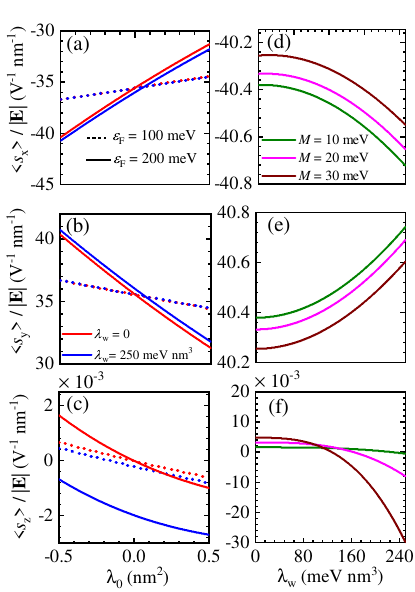}
\caption{(Left panel): Spin-density as a function of $\lambda_0$, for different Fermi energies and two values of the warping term. 
(Right panel): Spin-densities along various directions as a function of the warping term, $\lambda_{w}$, for different magnetization values when the Fermi energy is set to 200 meV. The magnitude of $\langle s_z \rangle$ is four orders of magnitude smaller than the in-plane spin-density values. We assume $\phi_E=\phi_m=\pi/4$.}
\label{fig-spin}
\end{figure} 
\section{Numerical Results and Discussion}
We present some numerical estimates for spin density and spin-orbit torque components. In our numerical calculations, we mostly consider the following values: $\phi_{E}=\phi_{m}=\pi/4$, $v_{\rm F} = 6\times{10^5}$ m/s, $n_i = 10^{10}$ cm$^{-2}$, $\lambda_w = 250$ meV nm$^{3}$, and $\varepsilon_{\rm F}=200$ meV, $M = 10$ meV, and $u_0 = 5$ eV nm$^2$; unless otherwise stated.

We find that spin-orbit scattering significantly renormalizes the Rashba-Edelstein effect in the magnetized topological insulator surface states. It can enhance or reduce the induced spin density, and hence the spin-orbit torque, depending on the sign of the spin-orbit scattering coefficient $\lambda_0$. This is shown in the left panel of Fig.~\ref{fig-spin} where the spin density along $\hat{x}$, $\hat{y}$ and $\hat{z}$ as a function of $\lambda_0$. We find that the {total} spin density is enhanced when $\lambda_0<0$ and reduced when $\lambda_0>0$. { Despite the out-of-plane spin density not always following this trend, this is still generally true as the out-of-plane spin density is 4 orders of magnitude smaller than the in-plane.} This spin density is enhanced for negative values of $\lambda_0$ because the coefficient $Z(k_{\rm F})$ will be larger. Spin-orbit scattering can renormalize the REE spin polarisation by $\approx15\%$, similarly it can change the polarisation for the out-of-plane case by $\approx40\%$. This result is important as the ultimate goal of SOT experiments is to enhance their efficiency and achieve rapid and energy-efficient magnetization switching. Hence, considering the effects of extrinsic spin-orbit scattering is an important factor for device design. Furthermore, we would like to note that the strength of the impurity-induced spin-orbit coupling $\lambda_0$ could be enhanced by the proximity effect with another material with a strong spin-orbit interaction~\cite{hu2015controllable}.

\begin{figure}
\centering
\includegraphics[width=\linewidth]{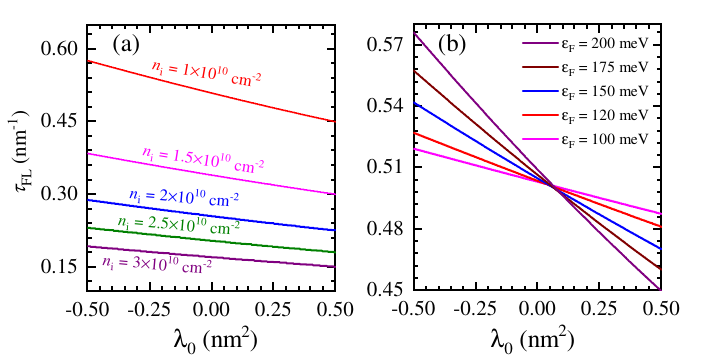}
\caption{{Field-like SOT as a function of the extrinsic SO scattering strength, $\lambda_0$ (a) for different values of impurity density and (b) for different values of the Fermi energy. Here, the damping-like SOT is not plotted as it is independent of impurity density. Other parameters are the same as those in Fig. \ref{compare}.}}
\label{fig-tauvsni}
\end{figure}

\begin{figure}
\centering
\includegraphics[width=\linewidth]{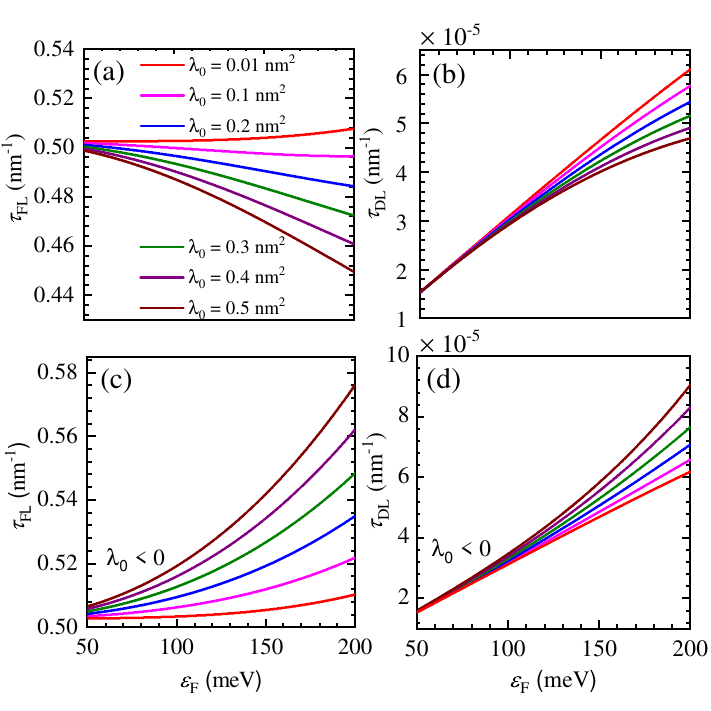}
\caption{{(Left panel): Field-like SOT as a function of the Fermi energy for various (a) positive and (c) negative values of the extrinsic SO scattering strength. 
(Right panel): Damping-like SOT as a function of the Fermi energy for various (b) positive and (d) negative values of the extrinsic SO scattering strength. Other parameters are the same as those in Fig. \ref{compare}.}}
\label{fig-tauvsef}
\end{figure}

{The magnitude of the enhancement or reduction of the torque density due to the spin-orbit scattering parameter increases with the Fermi energy, as is shown in Fig.~\ref{fig-tauvsni}. Furthermore, increasing the Fermi energy will increase or decrease the magnitude of the torque depending on the value of $\lambda_0$, as is shown in Fig.~\ref{fig-tauvsef}. The differing dependence of the torque on the Fermi energy due to $\lambda_0$ provides an experimental signature of the spin orbit-scattering parameter. Experimentally measuring the spin-orbit scattering parameter would be important since it cannot be calculated for topological insulators using the standard method for semiconductors. Similarly, the different dependencies of the out-of-plane and in-plane torques on $\lambda_0$, seen in Fig.~\ref{compare}, opens up another possibility for measuring its experimental signature, through measuring the difference in the magnitude of the field-like torque for and in-plane and out-of-plane magnetization. }

The leading term in the spin density (\ref{leading1}) is the dominant term, and is roughly 4 orders of magnitude greater than the other spin densities calculated for our chosen parameters. From this term we get the normal Rashba-Edelstien effect $\hat{\boldsymbol{z}}\times\boldsymbol{E}$ as well as a small spin polarisation out-of-plane driven by the hexagonal warping, as is shown in Fig \ref{fig-spin}, this is consistent with previous theoretical studies in topological insulators~\cite{Chang2015,Li2019,PhysRevB.96.014408}. Though the spin density contributions beyond the leading term are small, they do have some interesting properties. The intrinsic, extrinsic, side jump and skew scattering contributions to the spin density are independent of the scattering time. Whereas the leading term is linear in the scattering time and will be significantly reduced in disordered systems, these extra contributions will not be affected by disorder. However, we still expect the leading term to be dominant in highly disordered systems. {Hence, the spin torque will be greatly enhanced in cleaner systems as is shown in Fig.~\ref{fig-tauvsni}}. Despite being small, we expect the beyond leading contributions to be measurable due to their differing symmetry. Due to the interplay between the hexagonal warping and in-plane magnetization there are spin polarisations of the form $m_y\Tilde{\boldsymbol{E}}+m_x\hat{\boldsymbol{z}}\times\Tilde{\boldsymbol{E}}$, where $\Tilde{\boldsymbol{E}}=(E_x,-E_y,0)$. These spin polarisations will generate a non zero field-like torque for a electric field $\parallel\hat{x}$ and a magnetization $\parallel\hat{y}$. When this setup is rotated and the electric field is $\parallel\hat{y}$ and magnetization is $\parallel\hat{x}$ the field-like torque will be zero. Hence, it is possible to measure these extra torques experimentally.

The spin densities we calculated are of a similar order of magnitude to previous numerical studies~\cite{Ghosh2018,Chang2015}, and are in good agreement with similar studies of the REE in TI surface states employing the Kubo formula~\cite{Sakai_2018, PhysRevB.99.155139, PhysRevB.96.014408}. In addition to the normal REE spin density there is also an out-of-plane spin density driven by the warping terms, again we find our results to be in good agreement with previous works~\cite{PhysRevB.99.155139}. Though there have been a number of previous studies on topological insulator surface state spin torques, our work is distinct from them in some key areas. First, most of these works other than Ref.~\onlinecite{PhysRevB.99.155139} do not study the case where the magnetization is in plane. Second, they do not consider spin-orbit scattering effects, which we have shown are very important for the surface state spin torque. Furthermore, Ref.~\onlinecite{PhysRevB.99.155139} only gives results for the case where $\boldsymbol{E}\parallel\hat{x}$, due to warping removing the $x$-$y$ symmetry there are subtle differences in the spin torques generated for other electric field directions. We find that there will be a electrically induced spin density aligned $\parallel\boldsymbol{E}$ that will vanish when the electric field is aligned along $\hat{\boldsymbol{y}}$.

We comment on the differences in TI surface state spin-orbit torques between the cases with an in-plane and out-of-plane magnetization. As we have previously shown~\cite{farokhnezhad2022spin}, when the magnetization direction is fixed along the out-of-plane direction, the intrinsic and extrinsic contributions (such as skew and side-jump scatterings) induce a field-like SOT that is three orders of magnitude greater than the damping-like torque. In Fig.~\ref{compare} we plot the damping-like and field-like torques vs the spin-orbit scattering parameter for the case when the magnetization is in-plane and when it is out-of-plane. The ratio of field-like to damping-like torque for the in-plane case is 2 orders of magnitude larger than for the out-of-plane case. We find the spin-orbit scattering parameter has a similar relationship with the spin torque in both cases and can significantly enhance the spin torque. However, we find that the out-of-plane spin torque is more significantly affected by the spin-orbit scattering, as is demonstrated in Fig.~\ref{compare}. This opens up the possibility of measuring the  experimental signature of the spin-orbit scattering parameter, through measuring the difference in the magnitude of the field-like torque for an in-plane and out-of-plane magnetization. However, such a measurement may be limited due to the materials used in TI/FM devices with in-plane and out-of-plane magnetizations tending to be different, as different interfaces would likely result in different values of $\lambda_0$. Nevertheless, through careful engineering and material choice such a measurement should be possible. Furthermore, by engineering the interface and optimizing the thickness of the ferromagnetic layer, proximity effects can be enhanced, potentially allowing for stronger and hence more efficient spin-orbit torque. This work along with our previous work on surface state spin torques\cite{farokhnezhad2022spin} gives a complete picture of the surface state contribution to topological insulator spin torques. While there are a number of other mechanisms that contribute to the topological insulator SOT~\cite{Ghosh2018, cullen2023spin, cullen_2022, liu2023topological, PhysRevLett.121.136805, PhysRevB.94.104419, PhysRevB.94.104420, Kurebayashi2019, siu2018, edelstein1990spin, mihai2010current, mellnik2014spin, Dc2018}, it should be noted that the field-like torque due to the REE, renormalized by extrinsic spin-orbit scattering, is still expected to be the dominant contribution regardless of the sign of the extrinsic spin-orbit parameter $\lambda_0$.

\section{Conclusion}
Scattering due to spin-orbit coupled impurities significantly renormalizes the Rashba-Edelstein effect in the magnetized topological insulator surface states. Whether the spin density is reduced or enhanced depends on the chirality of the impurity potential. Whereas this renormalization can be as large as $\approx15\%$ for the case of topological insulator surface states with an in-plane magnetization, for the case of an out-of-plane magnetization it can be as large as $\approx40\%$. These results have significant implications for device design: the dominant field-like torque is determined by extrinsic spin-orbit scattering and can be an experimental signature of it.

\section{Acknowledgments}. This project is supported by Future Fellowship FT190100062. JHC acknowledges support from an Australian Government Research Training Program (RTP) Scholarship.


%

\end{document}